\documentclass[aps,prl,twocolumn,superscriptaddress, showpacs]{revtex4-1}
\usepackage{graphicx}
\usepackage{dcolumn}
\usepackage{bm}
\usepackage{lipsum}
\usepackage{amsmath}
\usepackage{epstopdf}
\begin{document}
\title {Unusual temperature evolution of superconductivity in LiFeAs}

\author{P. K. Nag}
\email{p.k.nag@ifw-dresden.de}
\author{R. Schlegel}
\author{D. Baumann}
\author{H.-J. Grafe}
\author{R. Beck}
\affiliation{Leibniz-Institute for Solid State and Materials Research, IFW-Dresden, 01069 Dresden, Germany}

\author{S. Wurmehl}
\affiliation{Leibniz-Institute for Solid State and Materials Research, IFW-Dresden, 01069 Dresden, Germany}
\affiliation{Institute for Solid State Physics, TU Dresden, 01069 Dresden} 

\author{B. B\"uchner}
\affiliation{Leibniz-Institute for Solid State and Materials Research, IFW-Dresden, 01069 Dresden, Germany}
\affiliation{Institute for Solid State Physics, TU Dresden, 01069 Dresden} 
\affiliation{Center for Transport and Devices, TU Dresden, 01069 Dresden, Germany}

\author{C. Hess}
\email{c.hess@ifw-dresden.de}
\affiliation{Leibniz-Institute for Solid State and Materials Research, IFW-Dresden, 01069 Dresden, Germany}
\affiliation{Center for Transport and Devices, TU Dresden, 01069 Dresden, Germany}

\date{\today}

\begin{abstract}

We have performed temperature dependent tunneling spectroscopy on an impurity-free surface area of a LiFeAs single crystal. Our data reveal a highly unusual temperature evolution of superconductivity: at $T_c^*=18$~K a partial superconducting gap opens, as is evidenced by  subtle, yet clear features in the tunneling spectra, i.e. particle-hole symmetric coherence peaks, and a dip-hump structure which signals strong-coupling superconductivity. At $T_c=16$~K, these features substantiate dramatically and become characteristic of full superconductivity. Remarkably, this is accompanied by an almost jump-like increase of the gap energy at $T_c$ to about 87\% of its low-temperature gap value. The energy of the bosonic mode as measured by the distance between the coherence peak and the higher-energy dip remains practically constant in the whole temperature regime $T\leq T_c^*$. The comparison of these findings with established experimental and theoretical results lead us to suggest that the bosonic mode is not directly related to incommensurate spin fluctuations that have previously been observed in inelastic neutron scattering.

\end{abstract}
\maketitle

The physics of the material LiFeAs seems to differ in many aspects from that of canonical iron-based superconductors, and accordingly attracts considerable attention. Unlike the latter, where superconductivity emerges from a Fermi surface-nested antiferromagnetic spin density wave (SDW) state upon doping \cite{Kamihara2008,Luetkens2009,Rotter2008,Sefat2008}, LiFeAs superconducts without any doping \cite{Tapp2008}. The electronic band structure of LiFeAs has been much under debate: after early de Haas-van Alphen (dHvA) and scanning tunneling spectroscopy (STS) experiments from which a nested Fermi surface had been deduced \cite{Putzke2012,Allan2012}, there exists now a comprehensive data set from transport \cite{Heyer2011}, dHvA \cite{Zeng2013}, STS \cite{Haenke2012,Hess2013}, and inelastic neutron scattering (INS) \cite{Qureshi2012,Qureshi2014} experiments which fully support \cite{Hess2013,Knolle2012} reports of a Fermi surface without any nesting from angular resolved photoemission spectroscopy (ARPES)  \cite{Borisenko2010,Kordyuk2011,Umezawa2012,Borisenko2012}. Strong controversy still exists about the nature of superconductivity: various order parameter symmetries, including singlet $s^{+-}$-wave and  $s^{++}$-wave as well as triplet $p$-wave states have been proposed in both theoretical and experimental work \cite{Brydon2011,Baek2012,Haenke2012,Platt2011,Borisenko2012,Chi2014,Wang2013,Ahn2014,Yin2014}. 

In this letter, we report a careful exploration of the temperature dependence of the superconducting state of LiFeAs as seen in scanning tunneling spectroscopy (STS) of a well-defined clean area of the LiFeAs surface. The most striking observation is an apparent phase transition within the superconducting state with dramatic impact on the spectral signatures of superconductivity in the differential conductance $dI/dV$. More specifically, we observe the onset of faint superconductivity at $T_c^*=18$~K as evidenced by a particle-hole symmetric depletion and coherence peaks seen in the $dI/dV$ at zero bias voltage and at about $\pm$3-4~mV, respectively. In addition, characteristic strong-coupling dip-hump modifications in the $dI/dV$ appear on both polarities at a distance of about 5-6~meV from the coherence peaks. These features affect only a tiny but well resolvable portion of the total $dI/dV$. Upon lowering the temperature ($T$), however, these features become dramatically enhanced at $T_c=16$~K and rapidly develop into those of a fully gapped superconducting state upon further cooling. At the transition at 16~K, the coherence peaks shift to about $\pm 6$~mV which is already about 87\% of the low-$T$ gap value. Remarkably, the energy of the coupled bosonic mode stays largely unaffected from this transition. We discuss this unusual non-BCS-like $T$-evolution of superconductivity in connection with recent ARPES, INS, and theory results.

In previous experimental works the reported critical temperature $T_c$ for stoichiometric LiFeAs scatters between about 15~K and 18~K \cite{Pitcher2010, Aswartham2011, Khim2011, Li2013, Stockert2011, Wang2008, Heyer2011, Pitcher2008, Chi2012, Hanaguri2012}. Interestingly the occurrence of multiple critical temperatures has previously been reported from successive nuclear magnetic resonance (NMR) Knight shift and AC-susceptibility measurements of one LiFeAs single crystal \cite{Baek2013}. However, in such an experiment, it is very difficult to rule out sample inhomogeneity which in principle could cause the mentioned probes to respond to different parts of the sample, each with a potentially different critical temperature. In order to exclude such thinkable complications, we took great care in the sample synthesis, its characterization and the STS measurements.

Single crystals of stoichiometric LiFeAs have been grown using the self-flux method as described in Ref.~\onlinecite{Morozov2010}. In order to confirm stoichiometry and homogeneity of the sample prior to mounting it into our scanning tunneling microscope (STM), we determined \cite{supplemental} the $^{75}$As nuclear quadrupole resonance (NQR) frequency and line width of the sample as 21.54~MHz and 24~kHz, respectively. While the frequency is consistent with previous findings, the measured line width is by a factor of 1.5-3 times sharper than that of previously investigated samples \cite{Qureshi2012,Baek2012,Baek2013, Morozov2010} highlighting the extraordinary homogeneity of our crystal. Since LiFeAs is highly air sensitive, these steps, and the actual mounting of the sample into our home-built STM \cite{Schlegel2014} have been performed in Ar atmosphere. Once the sample had been mounted into the STM, the sample space was evacuated, and cooled to about 5~K. Subsequently, the sample was cleaved in cryogenic vacuum just before approaching an electrochemically etched W-tip for performing the tunneling measurements. 
Topography measurements have been executed in constant-current mode, while for measuring the $dI/dV$ we used a lock-in amplifier with a modulation of 0.4~mV~rms at 1.1111~kHz. In the used sign convention of data, negative bias voltages probe the occupied states of the sample. In all shown spectroscopic data figures, an unavoidable offset of -0.6~mV has been corrected. 

During the $dI/dV$-measurements at different $T$ we ensured that the tunneling tip always probed the very same atoms within a 2~nm~$\times$~2~nm area to rule out any possible influence of sample inhomogeneity. Within this area, 800 grid spectra (including forward and backward sweep) have been recorded and averaged in the range $\pm$35 mV with 0.35 mV energy point resolution. Thereby, the system was stabilized at -35~mV before and after every single forward and backward sweep. At each $T$-level, we let the system sufficiently thermalize in order to have thermal drift less than an atom after every set of grid spectra, i.e. after about 2 hours. Furthermore, we confirmed an unchanged tip state before and after each grid spectroscopy from comparing corresponding topography scans \cite{supplemental}.

\begin{figure}[ht]
	\centering
		\includegraphics[width=\columnwidth]{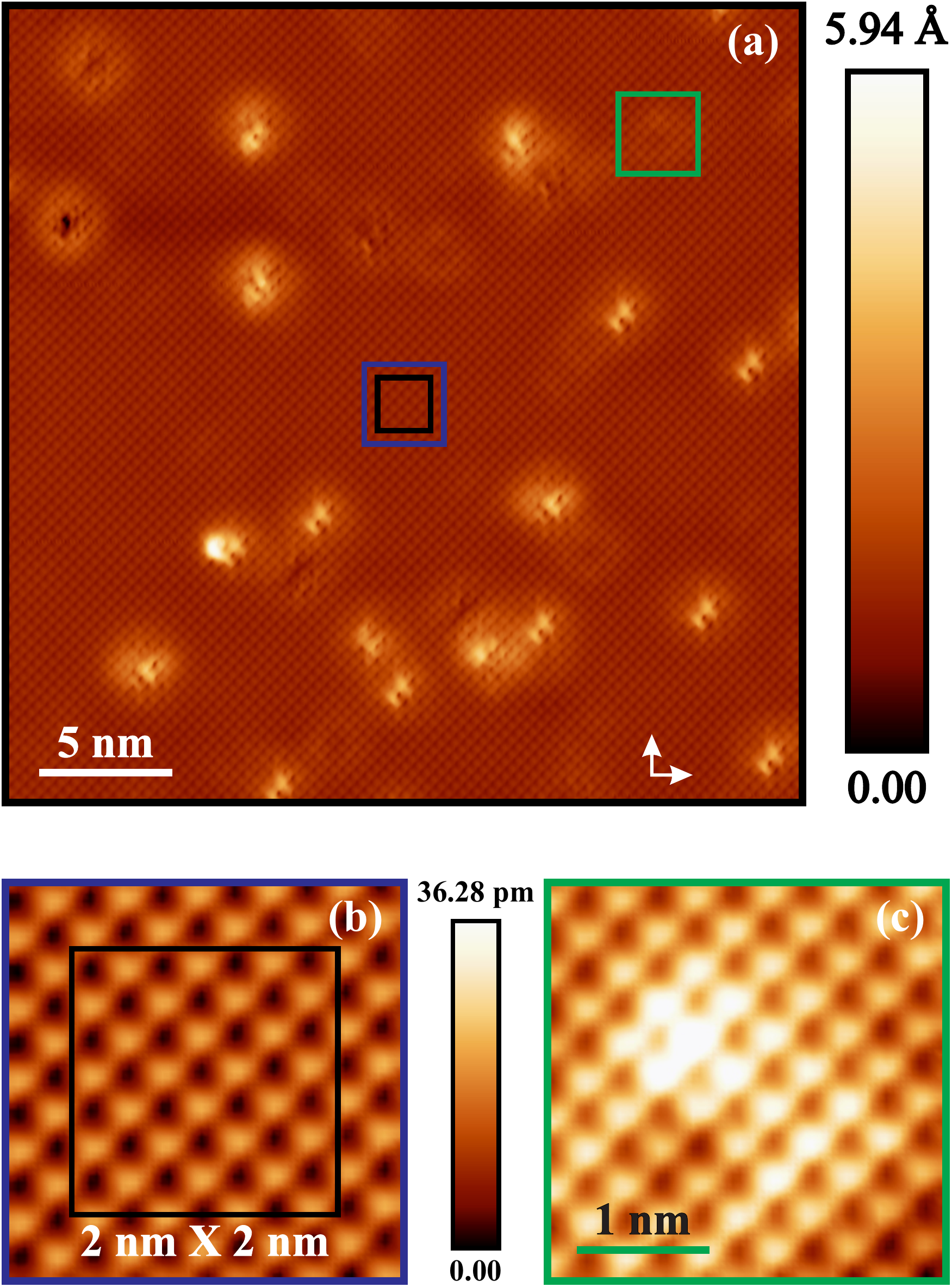}
\caption{(Color online) (a) 30 nm x 30 nm area of atomically resolved constant current mode topography image of LiFeAs ($\textit{I}_T$ = 300 pA, $\textit{V}_{bias}$ = +35 mV) measured at $T=4.8$~K. White arrows indicate the in-plane shortest Fe-Fe directions. The atomic corrugation on the surface corresponds to the Li-Li (As-As) lattice spacing of 3.77~\AA. 22 bright impurities from the first layer appear within the scan area. Faint signatures of impurities presumably of the second layer of the material are also visible (green square). Temperature dependent spectroscopy has been measured within the black square of 2 nm x 2 nm area. 
(b) Zoom-in into the blue square in Fig.~\ref{topoat_35_mV_Fe_direction_final}a to show atomic contrast in absence of impurities. (c) Zoom-in into the green square in Fig.~\ref{topoat_35_mV_Fe_direction_final}a to show the influence of an impurity in the second layer. }
\label{topoat_35_mV_Fe_direction_final}
\end{figure}

Fig.~\ref{topoat_35_mV_Fe_direction_final}a shows a representative topography scan with a field of view of 30 nm $\times$ 30 nm, measured at 4.8~K. The atomically resolved corrugation reveals the position of about 3000 surface Li-atoms \cite{Haenke2012} with a lattice spacing of roughly 3.77 \AA, matching the reported  Li-lattice constant of the material \cite{Tapp2008}. Isolated locations with bright contrast represent impurity states of the first layer of the material \cite{Grothe2012}. Beside these, faint structures appear in the image, presumably, arising from second  layer impurity states as highlighted in Fig.~\ref{topoat_35_mV_Fe_direction_final}c. In order to spectroscopically investigate the pristine electronic structure of the material we selected a 2~nm $\times$ 2~nm area (indicated by a black square area in Fig.~\ref{topoat_35_mV_Fe_direction_final}a and Fig.~\ref{topoat_35_mV_Fe_direction_final}b), far away from first and second layer impurity states, and performed scanning tunneling spectroscopy as a function of $T$ in this area.

\begin{figure}[ht]
	\centering
		\includegraphics[width=\columnwidth]{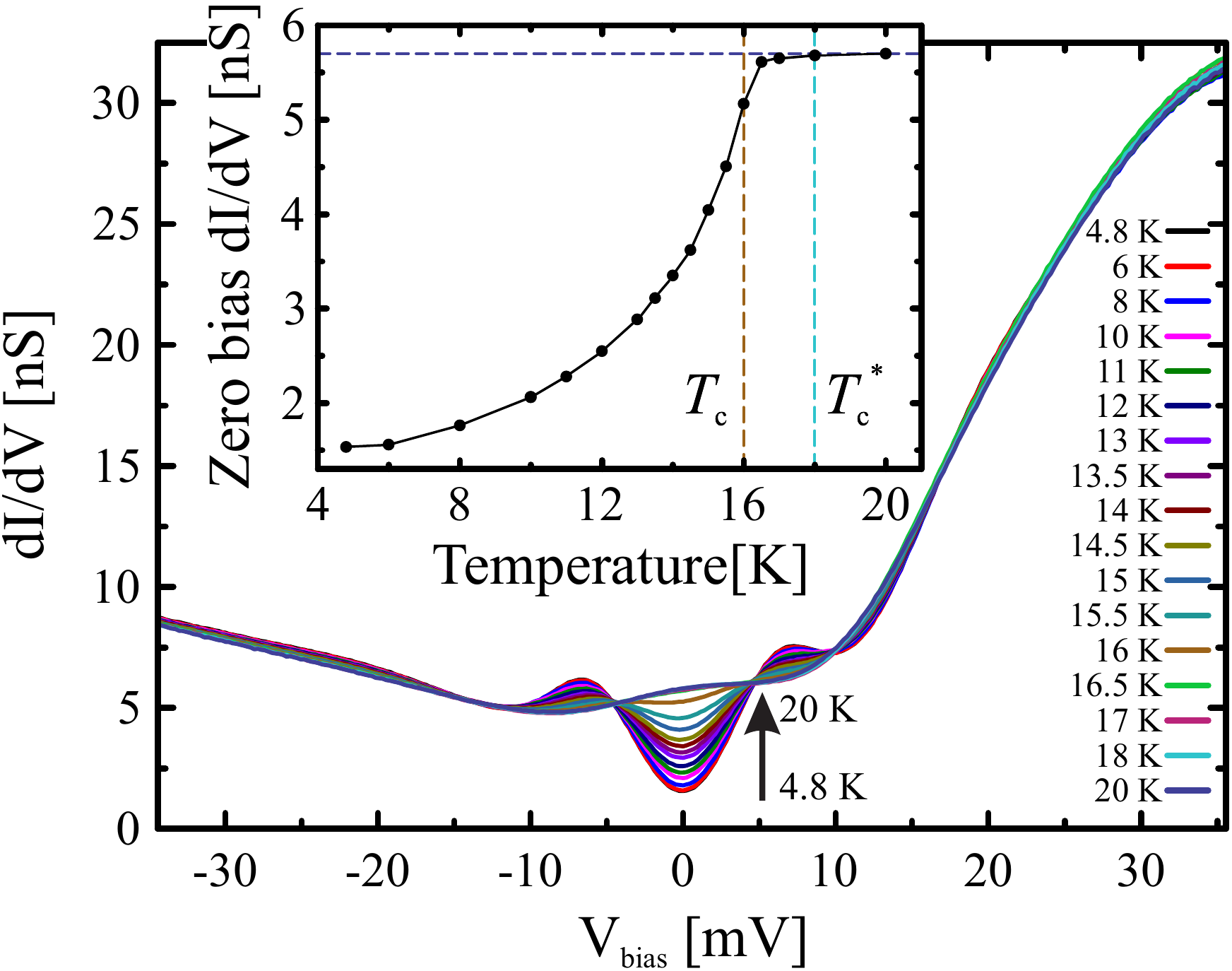}
\caption{(Color online) Temperature dependent tunneling spectra measured within the black square of Fig.~\ref{topoat_35_mV_Fe_direction_final}a/b between 4.8~K and 20~K. The up-arrow indicates the order of the curves at $V_\mathrm{bias}$ = 0 with increasing temperature. Inset: Zero bias differential conductance as a function of temperature. The horizontal dashed line is a guide to the eye. Vertical dashed lines indicate $T_c$ and $T_c^*$, see text.}
\label{raw_TDS}
\end{figure}

\begin{figure*}[ht]
	\centering
		\includegraphics[width=\textwidth]{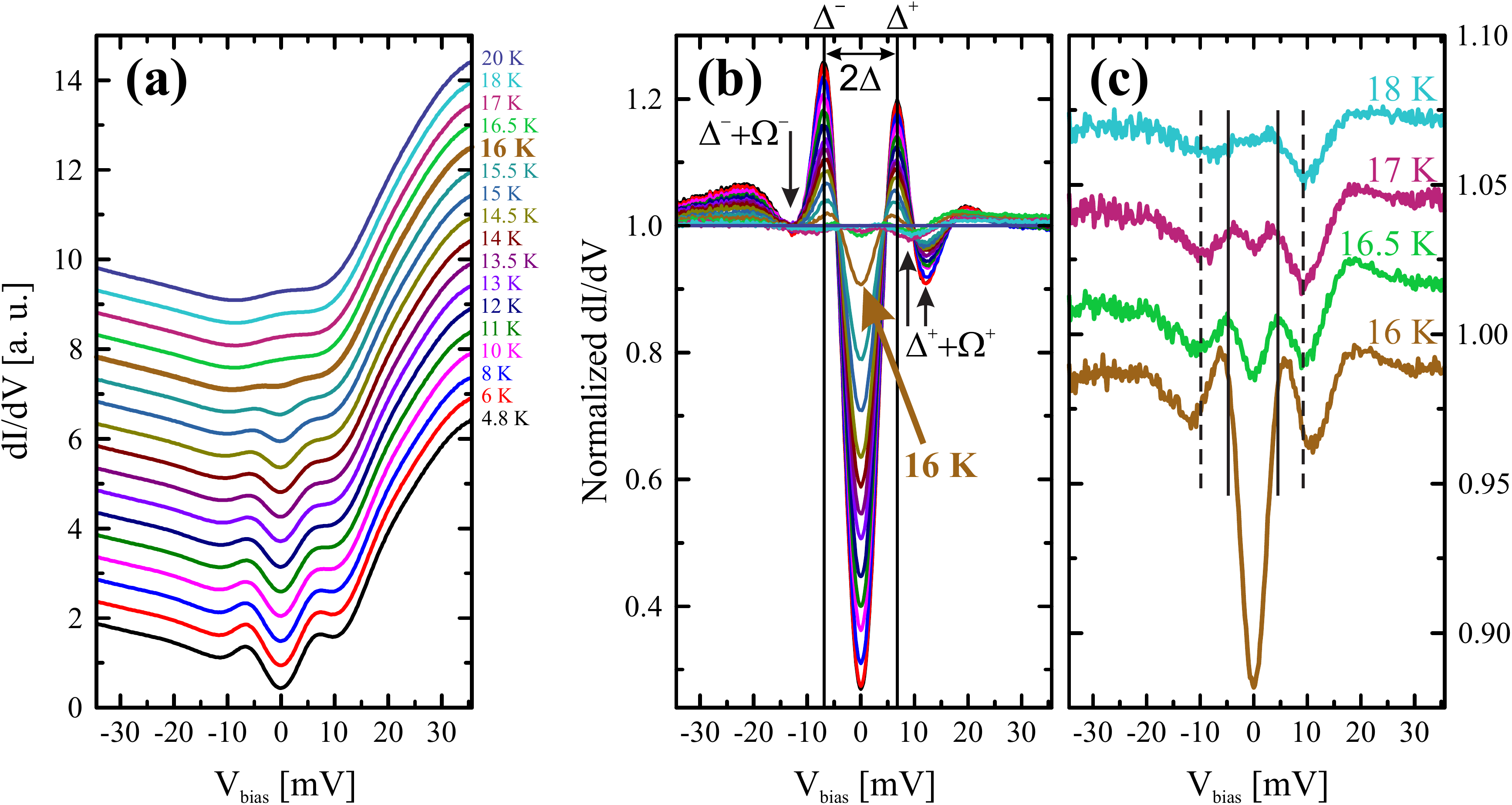}
\caption{(Color online) (a) Waterfall representation of the differential conductance $dI/dV$ for various temperatures. The spectrum at 16~K is highlighted in bold. (b) Differential conductance $dI/dV$ at various temperatures normalized to that at 20~K. Black up-arrows indicate the shift of the position of the positive energy dip at $\Delta^++\Omega^+$ towards lower energy upon raising the temperature through $T_c=16$~K. The down-arrow indicates the coarse position of the negative energy dip at $-\Delta^--\Omega^-$. (c) Waterfall representation of normalized spectra in b) at 16~K to 18~K. Superconducting coherence peaks and dip positions are indicated by solid and dashed vertical lines, respectively.}
\label{Temp_dependency_measurement}
\end{figure*}

Our results for the averaged differential conductance $dI/dV$ within the area of inset of Fig.~\ref{topoat_35_mV_Fe_direction_final}a as a function of applied bias voltage $V_\mathrm{bias}$ for the different $T$-levels are shown in Fig.~\ref{raw_TDS}. Two aspects can be recognized in the shown data:  i) in the normal state (20~K) the $dI/dV$ exhibits a strong asymmetry between occupied and unoccupied states with a hump-like enhancement around zero bias, in agreement with previous findings \cite{Hanaguri2012, Chi2012}. ii) At base temperature of our system (4.8~K) pronounced signatures of the superconducting state are superimposed on the normal state $dI/dV$ background, where the depletion of the $dI/dV$ at zero bias and the appearance of coherence peaks at finite bias voltages are the most prominent indicators of the superconducting gap. Upon increasing $T$, the data reveal a systematic closing of the gap with an apparent $T_c$ of about 16~K (see also Fig.~\ref{Temp_dependency_measurement}a). For illustrating the gap closing we plot the $dI/dV$ at the energy of strongest depletion close to zero bias in the inset of Fig.~\ref{raw_TDS}. The onset of a strong decrease of $dI/dV$ below 16~K is clearly visible. However, a close inspection of the data reveals that the closing of the gap seems incomplete even at temperatures $T>16$~K. 

In order to obtain further insight into the data, we normalize all tunneling spectra with respect to the normal state spectrum at 20~K, see Fig.~\ref{Temp_dependency_measurement}b. In this representation, all features of the superconducting state become better visible. At small energies within the gap (as marked by the position of the coherence peaks) the spectra become almost particle-hole symmetric. Remarkably, this symmetry does not persist at larger energies. In particular, at positive bias a pronounced dip and a hump are present in the spectra at all temperatures (see Fig.~\ref{Temp_dependency_measurement}b and Fig.~\ref{Temp_dependency_measurement}c). However, at negative bias, the dip corresponds to $dI/dV$ values barely below that of the normal state, whereas the hump structure is even more pronounced. Such dip-hump features are known as the signature of strong-coupling superconductivity \cite{Scalapino1966}, and are frequently observed in different Fe-based superconductors, where these are often interpreted as fingerprints of spin fluctuations \cite{Wang2013a,Chi2012,Shan2012,Song2014}. Surprisingly, the canonical spectral signatures of superconductivity, i.e. the coherence peaks, and the dip-hump structure are even clearly present in the normalized tunneling data \textit{above} the afore inferred $T_c$ of 16~K. This is unambiguously shown in Fig.~\ref{Temp_dependency_measurement}c which focuses on the $T$-range 16~K to 18~K. Thus, in addition to the onset of pronounced superconductivity at $T_c=16$~K, our LiFeAs sample is in a superconducting state that is characterized by faint corresponding spectral features already below $T_c^*=18$~K.

\begin{figure}[ht]
	\centering
		\includegraphics[width=\columnwidth]{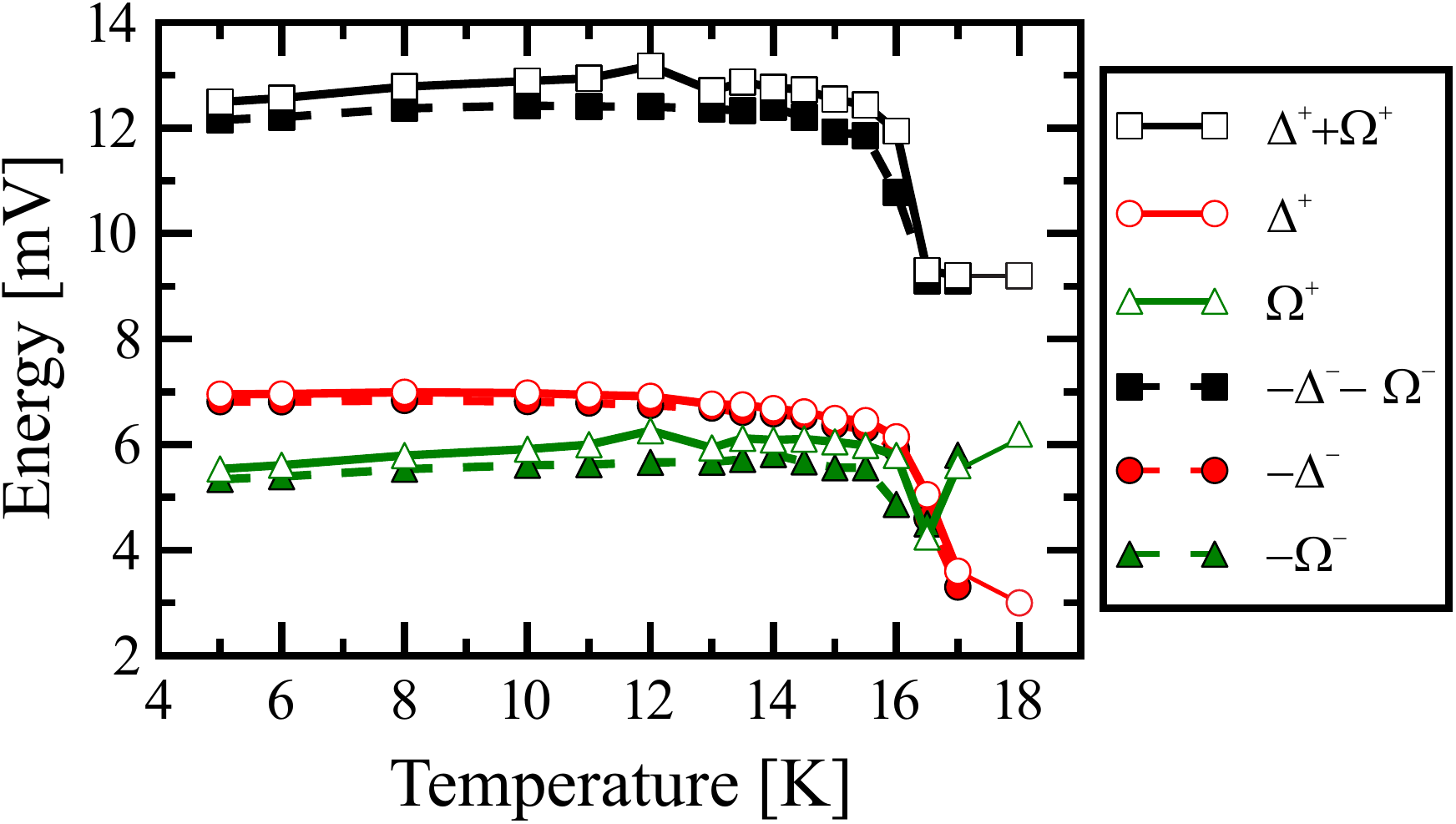}
\caption{(Color online) Temperature evolution of superconducting coherence peaks ($\Delta^+$, $-\Delta^-$), the dip positions ($\Delta^++\Omega^+$, $-\Delta^--\Omega^-$) and the resulting mode energies ($\Omega^+$ and $-\Omega^-$).}
\label{peak_and_dip_position}
\end{figure}

After having established this main experimental observation of this letter, we turn now to thoroughly analyzing the observed spectral features as a function of $T$. In Fig.~\ref{Temp_dependency_measurement}b we assign the distance between the coherence peaks at positive (negative) energy $\Delta^+$ ($\Delta^-$) to the double gap value $2\Delta$ and the distance between $\Delta^+$ ($\Delta^-$) and the dip position at its high energy side to the mode energy $\Omega^+$ ($\Omega^-$). At base temperature (4.8~K) we find $\Delta\approx$~6.9 mV consistent with previous findings \cite{Chi2012, Hanaguri2012}. For the mode energy we find $\Omega^+\approx |\Omega^-|=5.4\pm0.1$~mV \footnote{The coherence peaks and dip positions have been determined with a phenomenological parabolic fit to the corresponding extrema. The given errors are phenomenological estimates.}.
The $T$-evolution of superconducting coherence peaks, the dip positions, and the resulting mode energies are summarized in Fig.~\ref{peak_and_dip_position}. The peak positions which provide an estimate for the $T$-dependence of the superconducting gap $\Delta(T)$, interestingly, remain almost constant up to almost $T_c$. To be specific, $\Delta(T_c=16~\mathrm{K})/\Delta(4.8~\mathrm{K})\approx0.87$. Upon increasing $T$ further, $\Delta(T)$ drops abruptly to about 50\% of its low-$T$ value at 17~K and becomes barely resolvable at $T_c^*=18$~K.  This unusual behavior is in clear contrast to any BCS-like weak coupling scenario \cite{Tinkham} as has been previously suggested by Chi et al. \cite{Chi2012}. A very similar $T$-dependence is found for the positions of the dips. These remain almost constant at a value of about $\pm12$~mV up to $T_c=16$~K, and jump-like decrease to about $\pm9$~mV at higher $T$ up to $T_c^*=18$~K. Remarkably, the mode energy exhibits only a small dip at around $T_c$ but stays practically uninfluenced at lower and higher $T$.

The observation of two transition temperatures $T_c$ and $T_c^*$ at an atomically precisely defined microscopic position reveals these as an intrinsic property of the material. Thereby it corroborates earlier findings where the measured critical temperature on the very same sample depended on the probing method and offers a reconciliation of the spread of reported $T_c$ values \cite{Baek2013, Pitcher2010, Aswartham2011, Khim2011, Li2013, Stockert2011, Wang2008, Heyer2011, Pitcher2008, Chi2012, Hanaguri2012}.

A multiband electronic structure can in principle be pictured as a possible source for a complicated superconducting state with multiple possible order parameters \cite{Agterberg1999}. 
In LiFeAs, according to the ARPES-derived electronic band structure \cite{Wang2013a}, the Fermi surface consists of quasi two-dimensional hole-like (labelled $\gamma$) and electron-like pockets (labelled $\beta$) centered around the $\Gamma$- and $M$-points, respectively. Two further hole-like Fermi surface pockets (labelled $\alpha$) are centered around the $Z$-point. The latter are tiny, yet have been reported to possess the largest superconducting gap of about 6~meV as compared to about 3.5 to 4~meV at the $\gamma$- and $\beta$-pockets \cite{Umezawa2012,Borisenko2012}. Recent theoretical work which focuses on analyzing possible pairing scenarios on basis of this band structure suggests a complicated multigap scenario which allows for peculiar $T$-dependencies \cite{Ahn2014}. In particular, it has been proposed that the $\alpha$-Fermi surface pockets at $k_z\approx\pi$ may cause Cooper pairing prior to that of Fermi surface pockets at $k_z\approx0$, where the $\alpha$-bands remain below the Fermi level. As a consequence, at temperatures just below the onset of superconductivity, the superconducting state may be very different from that at lower $T$, where all Fermi surface pockets contribute to the superconductivity. Considering such a scenario, the observed $T_c^*=18$~K might be interpreted as the onset of superconductivity at $k_z\approx\pi$, whereas the critical temperature $T_c=16$~K can be related to the onset of full superconductivity in the complete Fermi surface at all $k_z$.

The second unusual observation in our data is the practically $T$-independent value of the mode energy $\Omega$, which implies that the dominant bosonic coupling mode remains the same for all $T$ below $T_c^*$. In view of the afore discussed scenario, this suggests that the observed mode primarily couples to the $\alpha$-bands, at all temperatures. Previously, the mode has been suggested \cite{Chi2012} to be connected with incommensurate spin fluctuations that have been observed in inelastic neutron scattering \cite{Qureshi2012,Qureshi2014}. However, the latter have been shown to be unrelated to particle-hole excitation of the $\alpha$-bands but instead emerge from those between the $\gamma$- and the $\beta$-bands \cite{Knolle2012, Qureshi2014}. Thus, the observed bosonic mode is likely to stem from another, yet to be discovered excitation in the system.

In conclusion, we have presented temperature dependent tunneling spectroscopy on the bare surface of stoichiometric LiFeAs with stable atomic registry. We observe a highly unusual temperature evolution of the superconducting state with the onset of faint superconductivity at $T_c^*=18$~K, which develops into full superconductivity only at $T_c=16$~K. In the whole superconducting temperature regime $T\leq T_c^*$ we observe evidence for a strong coupling bosonic mode that remains constant in energy except a slight reduction at $T_c$. The comparison of our findings with results from inelastic neutron scattering, experimental band structure data from ARPES, and theory let us to suggest that the bosonic mode might be unrelated to observed incommensurate spin fluctuations.

This project has been supported by the Deutsche Forschungsgemeinschaft through the Priority Programme SPP1458 (Grant HE3439/11 and BU887/15-1), and the Graduate School GRK1621.  Furthermore, this project has received funding from the European Research Council (ERC) under the European Union’s Horizon 2020 research and innovation programme (grant agreement No 647276 -- MARS -- ERC-2014-CoG). S.W. acknowledges funding by DFG under the Emmy-Noether program (Grant No. WU595/3-3). Discussions with P. Coleman, S. Borisenko, S.-H. Baek, A. Chubukov, and I. Eremin are gratefully acknowledged.

%

\end{document}